\documentclass[epj,nopacs]{svjour}
\usepackage{graphics}
\usepackage{epsfig}
\usepackage[english]{babel}
\usepackage{euscript,amstext,amssymb,amsbsy,fancybox,epsf}

\title{The Fubini-Furlan-Rossetti Sum Rule Revisited}
\author{B. Pasquini\inst{1}, D. Drechsel\inst{2}, L. Tiator\inst{2}
}                     

\authorrunning{B. Pasquini et al.}%
\institute{Dipartimento di Fisica Nucleare e Teorica, Universit\'a degli Studi
di Pavia; INFN, Sezione di Pavia, Pavia, Italy, and ECT*, Villazzano (Trento),
Italy, \and Institut f\"ur Kernphysik, Johannes Gutenberg-Universit\"at,
D-55099 Mainz }
\date{Received: date / Revised version: date}
\begin{document}

\def\dsdt{$\frac{d\sigma}{dt}$}
\def\beqn{\begin{eqnarray}}
\def\eeqn{\end{eqnarray}}
\def\barr{\begin{array}}
\def\earr{\end{array}}
\def\btab{\begin{tabular}}
\def\etab{\end{tabular}}
\def\bite{\begin{itemize}}
\def\eite{\end{itemize}}
\def\bcen{\begin{center}}
\def\ecen{\end{center}}

\def\eq{\begin{equation}}
\def\ee{\end{equation}}
\def\eqa{\begin{eqnarray}}
\def\eea{\end{eqnarray}}

\def\sl#1{\slash{\hspace{-0.2 truecm}#1}}

\abstract { The Fubini-Furlan-Rossetti sum rule for pion photoproduction on the
nucleon is evaluated by dispersion relations at constant $t$, and the
corrections to the sum rule due to the finite pion mass are calculated. Near
threshold these corrections turn out to be large due to pion-loop effects,
whereas the sum rule value is closely approached if the dispersion integrals
are evaluated for sub-threshold kinematics. This extension to the unphysical
region provides a unique framework to determine the low-energy constants of
chiral perturbation theory by global properties of the excitation spectrum.
} 
\maketitle

\section{Introduction \label{sec:intro}}

The Fubini-Furlan-Rossetti (FFR) sum rule was derived  on the basis of current
algebra and PCAC in the soft-pion limit~\cite{Fub65}. It relates the anomalous
magnetic moment to single-pion photoproduction on the nucleon. By use of the
Goldberger-Treiman relation~\cite{Gol58} the sum rule takes the form
\begin{equation}
\kappa^{V,S} = \frac{8M_N^2}{e\pi g_{\pi N}}\int\frac{d\nu'}{\nu'}\,
{\rm{Im}}\,A_1^{(+,0)}(\nu',\,t=0)\,,
\label{eq:intro}
\end{equation}
with $\kappa^{V} = \kappa_p-\kappa_n=3.706$ and  $\kappa^{S} =
\kappa_p+\kappa_n=-0.120$ the isovector and isoscalar anomalous magnetic
moments, and $A_1^{(+,0)}$ the respective combinations of the first invariant
amplitude of pion photoproduction. Furthermore, $M_N$ is the nucleon mass,
$M_\pi$ the mass of the neutral pion, and $g_{\pi N}$ the pseudoscalar
pion-nucleon coupling constant. The imaginary part of the amplitude
$A_1^{(+,0)}$ is evaluated at small momentum transfer $t$, and the
crossing-symmetrical variable $\nu'$ approaches the photon lab energy in the
soft-pion limit ($M_\pi\to 0,\,t\to 0$). In their derivation the authors of
Ref.~\cite{Fub65} tacitly assume that the RHS of Eq.~(\ref{eq:intro}) is
evaluated in the chiral limit of massless pions. However, the finite mass
corrections contained in the experimental data for ${\rm{Im}}\,A_1^{(+,0)}$
will yield deviations from the sum rule value, even though the suggested path
of integration at $t=0$ (corresponding to forward production of massless pions)
is expected to minimize these corrections.

The isovector sum rule was found to be slightly over-predicted in 1966 by a
then existing model of $\Delta(1232)$ resonance excitation~\cite{Fub66}. Adler
and Gilman~\cite{Adl66} generalized the sum rule to pion electroproduction and
evaluated the RHS of Eq.~(\ref{eq:intro}) with an early multipole
analysis~\cite{Sch64}. While the $\Delta(1232)$ multipoles yielded only about
60\% of the sum rule value, the non-resonant S-wave multipoles contributed
another 25\%. The authors of Ref.~\cite{Adl66} derived the sum rule in the
limit of vanishing values for $\nu,\,\nu_B$, and $M_\pi^2$, where
$\nu_B=(t-M_\pi^2)/4M_N$ and $\nu=\pm\nu_B$ gives the position of the nucleon
poles. The soft pion limit was approached by first letting $\nu_B\to 0$ and
then setting $\nu=0$, and the lower limit of the integral was fixed at
$\nu'=\nu_B+M_\pi+M_\pi^2/(2M_N)$.

In a more recent investigation Arndt and Workman~\cite{Arn95} used the VPI data
basis and obtained the values 3.92 and $-0.138$ from the RHS of
Eq.~(\ref{eq:intro}) for $\kappa^V$ and $\kappa^S$, respectively.

The FFR sum rule relies on two basic assumptions. First, the amplitudes
$A_1^{(+,0)}$ have to be evaluated at the origin of the Mandelstam plane,
$\nu=\nu_B=t=0$, which also requires that the pion mass vanishes. In this point
of the Mandelstam plane the pole terms due to the Dirac current vanish, and the
only contribution stems from the Pauli current, which yields a constant
(non-pole) contribution proportional to the anomalous magnetic moment (LHS of
Eq.~(\ref{eq:intro})). It is thereby assumed that all terms beyond the nucleon
pole graphs vanish at $\nu=t=M_\pi=0$. The second assumption concerns the
integral on the RHS of Eq.~(\ref{eq:intro}), which is evaluated by means of
dispersion relations at $t=const.$ In order to describe the amplitude at
$\nu=t=0$, the imaginary part under the integral therefore should be evaluated
at $t=0$, which for $M_\pi\ne0$ is outside the physical region of the
Mandelstam plane. The closest approximation to $t=0$ is forward pion
production, $\theta=0$, which then requires an extrapolation of the amplitude
from $t(\nu,\theta)<0$ to $t=0$.

\begin{figure}[ht]
\resizebox{0.47\textwidth}{!}{%
  \includegraphics{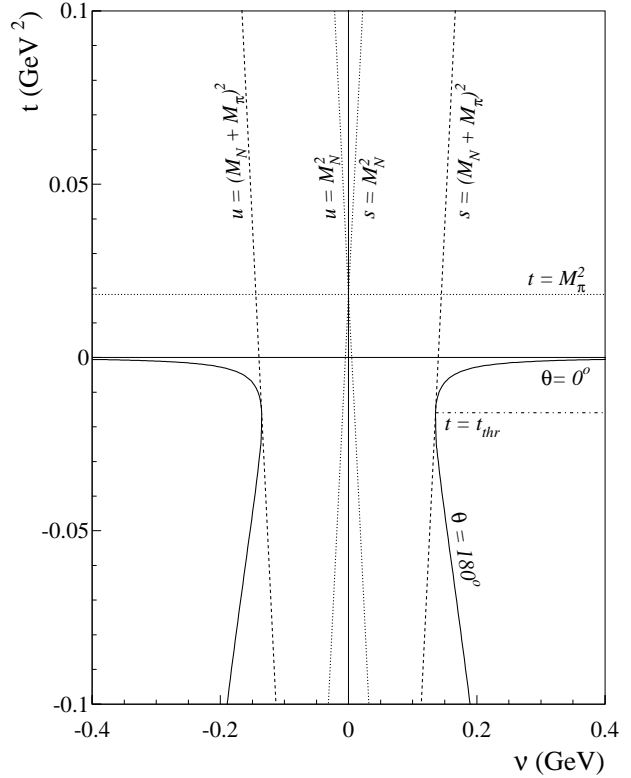}
}
\caption{The Mandelstam plane for pion photoproduction on the nucleon. The
boundaries of the physical region are $\theta=0$ (forward production) and
$\theta=180^{\circ}$ (backward production). The nucleon and pion pole positions
are indicated by the dotted lines $s=M^2_N$, $u=M_N^2$, and $t=M_\pi^2$. The
dashed line $s=(M_N+M_\pi)^2$ indicates the threshold of pion production and
therefore also of the imaginary part of the production amplitude. It is tangent
to the boundary of the physical region in the point $\nu=\nu_{\rm{thr}}$,
$t=t_{\rm{thr}}$. The path of integration starting in that point (dashed-dotted
line) yields the only dispersion relation at $t=const$ whose imaginary part is
fully contained in the physical region.
\label{mandelstam}}
\end{figure}

It is the aim of this work to investigate the FFR sum rule at $t=t_{\rm{thr}}$,
which yields the only path of integration that is completely within the
physical region (see Fig.~\ref{mandelstam}). The salient features of both
kinematics and photoproduction amplitudes are given in the following
section~\ref{sec:PPA}. The extrapolation to the soft-pion limit ($M_\pi\to 0$)
requires, of course, a dynamical framework. For this purpose we outline the
predictions of heavy baryon chiral perturbation theory (HBChPT) in
section~\ref{sec:HBChPT}. These predictions are compared to the results of
dispersion relations (DRs) based on MAID03 in section~\ref{sec:results}. We
present our results for the dispersion integrals at $t=t_{\rm{thr}}$ as
function of $\nu$, which near threshold ($\nu=\nu_{\rm{thr}}$) yield large cusp
effects from loop corrections but decrease rapidly with decreasing $\nu$ and
pass through a zero at $\nu\approx70$~MeV. A comparison of our approach with
HBChPT shows good agreement in the threshold region. However, the
nonrelativistic approximations of HBChPT turn out to be problematic in the
unphysical region far below pion threshold. Our findings confirm the necessity
to provide a fully relativistic treatment of ChPT, and at the same time they
yield a framework to determine the low-energy constants (LECs) of such a theory
by global properties of the nucleon's excitation spectrum. In
section~\ref{sec:sum}  we close with a short summary and an outlook.

\section{Pion Photoproduction Amplitudes \label{sec:PPA}}

Let us first define the kinematics of pion photoproduction on a nucleon, the
reaction
\[
\gamma (k) + N(p_i)\to\pi(q) + N'(p_f)\ ,
\]
where the variables in brackets denote the four-momenta of the participating
particles. The familiar Mandelstam variables are
\begin{eqnarray}
s=(p_i+k)^2,\quad t=(q-k)^2,\quad u=(p_i-q)^2,
\end{eqnarray}
and
\begin{eqnarray}
\nu=(s-u)/4M_N
\end{eqnarray}
is the crossing symmetrical variable. This variable is related to the photon
lab energy $E_\gamma^{lab}$ by
\begin{eqnarray}
\nu=E_\gamma^{lab} + \frac{t-M_\pi^2}{4M_N}\,.
\end{eqnarray}
The physical s-channel region is shown in Fig.~\ref{mandelstam}. Its upper and
lower boundaries are given by the scattering angles $\theta=0$ and
$\theta=180^{\circ}$, respectively. The nucleon and pion poles lie in the
unphysical region and are indicated by the dotted lines at $\nu_s=\nu_B$
(s-channel) and $\nu_u=-\nu_B$ (u-channel), where
\begin{eqnarray}
\nu_B =\frac{t-M_\pi^2}{4M_N}\,.
\end{eqnarray}
The threshold for pion photoproduction lies at
\begin{eqnarray}
\nu_{{\rm thr}}
&=& \frac{M_{\pi} (2M_N+M_\pi)^2}{4M_N (M_N+M_\pi)}\ ,\nonumber\\
t_{{\rm thr}} &=& -\frac{M_{\pi}^2 M_N}{M_N+M_\pi}\,.
\end{eqnarray}

In the pion-nucleon center-of-mass (c.m.) system, we have
\begin{equation}
\begin{array}{rllrll}
p_i^\mu & = & (E_i, -\vec k), & p_f^\mu & = & (E_f, -\vec q),\nonumber\\
k^\mu & = & (|\vec k|, \vec k), &  q^\mu & = & (\omega, \vec q),
\end{array}
\end{equation}
where
\begin{eqnarray}
k&=&|\vec k|=\frac{s-M_N^2}{2\sqrt{s}},\quad
\omega=\frac{s+M_\pi^2-M_N^2}{2\sqrt{s}},\nonumber\\
q & = & |\vec q|=\left[\left(\frac{s+M_\pi^2-M_N^2}{2\sqrt{s}}\right)^2
-M_\pi^2\right]^{1/2}\nonumber \\
& = &\left[\left(\frac{s-M_\pi^2+M_N^2}{2\sqrt{s}}\right)^2
-M_N^2\right]^{1/2},\nonumber\\
E_i & = & W-k=\frac{s+M_N^2}{2\sqrt{s}} \nonumber \\
E_f & = &W-\omega=\frac{s+M_N^2-M_\pi^2}{2\sqrt{s}},
\end{eqnarray}
with $W=\sqrt{s}$ the c.m. energy.

The nucleon electromagnetic current can be expressed in terms of 4 invariant
amplitudes $A_i$~\cite{Che57,Han98},
\begin{eqnarray}
J^\mu = \sum_i A_i(\nu,t)\, M^\mu_i,
\end{eqnarray}
with the four-vectors $M^\mu_i$ given by
\begin{eqnarray}
M^\mu_1&=&
-\frac{1}{2}i\gamma_5\left(\gamma^\mu\sl{k}-\sl{k}\gamma^\mu\right)\, ,
\nonumber\\
M^\mu_2&=&2i\gamma_5\left(P^\mu\, k\cdot q-
q^\mu\,k\cdot P\right)\, ,\nonumber\\
M^\mu_3&=&-i\gamma_5\left(\gamma^\mu\, k\cdot q
-\sl{k}q^\mu\right)\, ,\nonumber\\\
M^\mu_4&=&-2i\gamma_5\left(\gamma^\mu\, k\cdot P
-\sl{k}P^\mu\right)-2M_N \, M^\mu_1\, ,
\label{eq:tensor}
\end{eqnarray}
where $P^\mu=(p_i^\mu+p_f^\mu)/2$ and the gamma matrices are defined as in
Ref.~\cite{Bjo65}.

The invariant amplitudes $A_i$ can be further decomposed into three isospin
channels ($a=1,2,3$),
\begin{eqnarray}
A_i^a=A_i^{(-)}i\epsilon^{a3b}\tau^b+A_i^{(0)}\tau^a+A_i^{(+)}\delta_{a3},
\end{eqnarray}
where $\tau^a$ are the Pauli matrices in isospace. The physical photoproduction
amplitudes are then obtained from the following linear combinations:
\begin{eqnarray}
A_i(\gamma p\rightarrow n\pi^+)&=&\sqrt{2}(A_i^{(-)}+A_i^{(0)}),\nonumber\\
A_i(\gamma p\rightarrow p\pi^0)&=&A_i^{(+)}+A_i^{(0)},\nonumber\\
A_i(\gamma n\rightarrow p\pi^-)&=&-\sqrt{2}(A_i^{(-)}-A_i^{(0)}),\nonumber\\
A_i(\gamma n\rightarrow n\pi^0)&=&A_i^{(+)}-A_i^{(0)}.
\label{linear_comb}
\end{eqnarray}
The FFR sum rule is derived from the pion-photoproduction amplitude in the
limit of $q^\mu \rightarrow 0$. As we note from Eq.~(\ref{eq:tensor}), the
four-vectors $M^\mu_2,$ $M^\mu_3,$ and $M^\mu_4$ vanish, and only the
four-vector $M^\mu_1$ survives in that limit.

The isospin amplitudes $A_1^I$ satisfy the following dispersion relation at
fixed~$t$:
\begin{eqnarray}
\label{eq:dr1}
&&{\rm Re} A^{(+,0)}_1(\nu,t)= \\
&&A_1^{(+,0)\,pole}(\nu,t) +\frac{2}{\pi}{\cal P}\int_{\nu_{thr}}^{\infty}{\rm
d}\nu'
\frac{\nu'\,{\rm Im}A_1^{(+,0)}(\nu',t)}{\nu'^2-\nu^2}\,,\nonumber\\
&&{\rm Re} A^{(-)}_1(\nu,t)= \\
&&A_1^{(-)\,pole}(\nu,t) +\frac{2\nu}{\pi}{\cal P}\int_{\nu_{thr}}^{\infty}{\rm
d}\nu' \frac{{\rm Im}A_1^{(-)}(\nu',t)}{\nu'^2-\nu^2}\,. \nonumber
\label{eq:dr2}
\end{eqnarray}
Similar relations may be obtained for the amplitudes
$A_2-A_4$~\cite{Bal61,Han98}.

The nucleon pole contributions $A_i^{I,pole}$ ($I=0,+,-$) are  given by
\begin{eqnarray}
A_1^{I,pole} & = & \ \ \ \frac{eg_{\pi N}}{2}
\left(\frac{1}{s-M_N^2}+\frac{\epsilon^I}{u-M_N^2}\right)\,,\nonumber \\
A_2^{I,pole} & = & -\frac{eg_{\pi N}}{t-m^2_\pi}
\left(\frac{1}{s-M_N^2}+\frac{\epsilon^I}{u-M_N^2}\right)\,,\nonumber \\
A_3^{I,pole} & = & -\frac{eg_{\pi N}}{2m_N}\frac{\kappa^{I}}{2}
\left(\frac{1}{s-M_N^2}-\frac{\epsilon^I}{u-M_N^2}\right)\,,\nonumber \\
A_4^{I,pole} & = & -\frac{eg_{\pi N}}{2m_N}\frac{\kappa^{I}}{2}
\left(\frac{1}{s-M_N^2}+\frac{\epsilon^I}{u-M_N^2}\right)\,,
\label{eq:a1-4pole}
\end{eqnarray}
with $\epsilon^+=\epsilon^0=-\epsilon^-=1$, $\kappa^{(+,-)}=
\kappa_p-\kappa_n$, and $\kappa^{(0)}=\kappa_p+\kappa_n$, where $\kappa_p$ and
$\kappa_n$ are the anomalous magnetic moments of the proton and the neutron,
respectively. The nucleon pole contributions are most easily constructed by
evaluating the tree-level diagrams with the pseudoscalar pion-nucleon coupling.
We note in particular that the amplitudes $A_1^{I,pole}$ and $A_2^{I,pole}$ are
independent of the anomalous magnetic moment $\kappa_N$ of the respective
nucleon, whereas the amplitudes $A_3^{I,pole}$ and $A_4^{I,pole}$ are
proportional to $\kappa_N$. In order to obtain the proper chiral structure, the
tree-level contribution is calculated with pseudovector (PV) coupling, and the
amplitudes $A_1^{(+,0)}$ change according to
\begin{equation}
A_1^{(+,0)_{\rm{PV}}} = A_1^{(+,0)_{\rm{pole}}}+ A_1^{(+,0)_{\rm{FFR}}}\,,
\label{eq:A1PV}
\end{equation}
where
\begin{equation}
A_1^{(+,0)_{\rm{FFR}}} = \frac{eg_{\pi N}\,\kappa^{(+,0)}}{4M_N^2}\,.
\label{eq:A1FFR}
\end{equation}
All other amplitudes remain unchanged.

The covariant amplitude $A_1$ can be expressed by the CGLN
amplitudes~\cite{Che57,Han98} $\mathcal{F}_i(i=1\ldots 4)$ as follows:
\begin{eqnarray}
\label{eq:F1_4}
\lefteqn{ A_1 =  \frac{4\pi}{\sqrt{(E_i+M_N)\,(E_f+M_N)}} } \vspace{0.3cm} \\
 &&\left\{
\frac{W+M_N}{W-M_N}\,\mathcal{F}_1 - (E_f+M_N)\,\frac{\mathcal{F}_2}{q}\right.
\nonumber \\ && \left. +
\frac{M_N(t-M_\pi^2)}{(W-M_N)^2}\,\frac{\mathcal{F}_3}{q} +
\frac{M_N(E_f+M_N)\,(t-M_\pi^2)}{W^2-M_N^2}\,\frac{\mathcal{F}_4}{q^2}
\right\}\ ,\nonumber
\end{eqnarray}
where $q=|\vec{q}|$ and all variables are expressed in the c.m. frame. Below
the $\Delta(1232)$ resonance, we may limit ourselves to the S-wave multipole
and to the three P-wave multipoles $E_{1+},\ M_{1+}$, and $M_{1-}$. In this
approximation, the CGLN amplitudes take the form
\begin{equation}
\begin{array}{rllrll}
\mathcal{F}_1 & \to & E_{0^+} + 3(M_{1^+} + E_{1^+})\cos\theta\,,
\\ \\
\mathcal{F}_2/q & \to & (2M_{1^+}+M_{1^-})/q\ ,
\\ \\
\mathcal{F}_3/q & \to & 3(E_{1^+}-M_{1^+})/q\ , \quad & \mathcal{F}_4& \to &0 \
,
\end{array}
\end{equation}
where $\theta$ is the c.m. scattering angle, which is related to the Mandelstam
variables by
\begin{equation}
\cos\theta=\frac{(s-M_N^2)^2-M_\pi^2(s+M_N^2)+2\,s\,t}{2\,q\,\sqrt{s} \,
(s-M_N^2)} \ .
\end{equation}
The P-wave contributions are often expressed by the three combinations
\begin{eqnarray}
{P}_1 & = & 3E_{1+}+M_{1+} - M_{1-}\ ,  \nonumber \\
{P}_2 & = & 3E_{1+}-M_{1+} + M_{1-}\ ,\nonumber \\
{P}_3 & = & 2M_{1+} + M_{1-}\ .
\end{eqnarray}
With these definitions the multipole expansion of Eq.~(\ref{eq:F1_4}) can be
cast into the form
\begin{eqnarray}
\label{eq:19}
&& A_1  =  \frac{4\pi\,(W+M_N)}{\sqrt{(E_i+M_N)\,(E_f+M_N)}\,(W-M_N)} \\
&& \left\{ E_{0+} +\left( \omega + \frac{W(t-M^2_\pi)}{W^2-M_N^2}\right
)\,\bar{P}_1
\right. \nonumber \\
&& \left. + \frac{M_N(t-M_\pi^2)}{W^2-M_N^2}\,\bar{P}_2 +
\frac{t}{W+M_N}\,\bar{P}_3 + \ldots \right\}\ , \nonumber
\end{eqnarray}
with $\bar{P}_i = P_i/q$ and the ellipses denoting the higher partial waves.

 The FFR sum rule follows in the limit $q^\mu\to0$, which we approach by
first going to the production threshold ($\vec{q}=0$ in the c.m. frame) and
then letting $M_\pi\to0$. In the limit $q=|\vec{q}|\to 0$, only the S-wave
multipole $E_{0^+}$, the slopes of the P-wave multipoles, and the curvatures of
the D-wave multipoles contribute to Eq.~(\ref{eq:19}). Furthermore, the
kinematical factors simplify at threshold, and Eq.~(\ref{eq:19}) takes the form
\begin{eqnarray}
\label{eq:a1_thr}
&&A_1(\nu_{thr}, t_{thr})=\frac{4\pi}{M_\pi}\sqrt{\frac{M_N+M_\pi}{M_N}}\\
&& \Big\{ E_{0+}-\frac{M_NM_\pi}{M_N+M_\pi}\,\bar{P}_2 -
\frac{M_NM_\pi^2}{(2M_N+M_\pi)(M_N+M_\pi)}\, \nonumber \\ && \left (\bar{P}_3
+6M_N\,\bar{D} \right )\Big\}\, ,\nonumber
\end{eqnarray}
where $\bar{D}=(M_{2+}-E_{2+}-M_{2-}-E_{2-})/q^2$, and all the multipoles have
to be evaluated at $q=0$. In particular we note that the amplitude $\bar{P}_1$
does not appear in Eq.~(\ref{eq:a1_thr}), because its kinematical prefactor
vanishes at threshold.

With these definitions it is straightforward to obtain the threshold value of
the invariant amplitude, Eq.~(\ref{eq:a1_thr}), from current multipole analyses
or chiral perturbation theory (ChPT). The second step involved in the FFR, the
extrapolation to $M_\pi=0$, can of course only be performed within a
theoretical framework like ChPT.

\section{Predictions of HBChPT \label{sec:HBChPT}}

The Born terms in ChPT are evaluated with pseudovector pion-nucleon coupling in
order to obtain the correct chiral threshold behavior. The additional loop and
counterterm contributions to the S- and P-wave amplitudes have been predicted
by HBChPT to fourth order in $1/M_N$~\cite{Ber96,BKM96,Ber01}. After
subtraction of the nucleon-pole terms given in Appendix~A, we find the
following amplitude for neutral pion photoproduction on a nucleon $N$:
\begin{eqnarray}
\lefteqn{A_1(\nu,t) -A_1^{pole}(\nu,t) =} \vspace{1cm} \nonumber \\ &&
\frac{e\, g_{\pi N}}{2M_N^2}\kappa_N\,\tau_3 + A_1^{{\rm loop}}(\nu,t)
+A_1^{{\rm ct}}(\nu,t)\,,
\label{eq:a1_nonpole}
\end{eqnarray}
where use has been made of Eqs.~(\ref{linear_comb}), (\ref{eq:A1PV}), and
(\ref{eq:A1FFR}). In Eq.~(\ref{eq:a1_nonpole}), $\kappa_N\, \tau_3$ takes the
values 1.793 for the proton and 1.913 for the neutron. The appearance of the
anomalous magnetic moment $\kappa_N$ in this expression is, of course, the
essence of the FFR sum rule.

Because the FFR term of Eq.~(\ref{eq:A1FFR}) is a constant, the associated
current contributes only to the partial waves $E_{0+}$ and $M_{1-}$,
\newline
\begin{eqnarray}
\label{eq:E0+_M1-}
\lefteqn{E_{0+}^{\rm FFR}  =  \frac{eg_{\pi N}\kappa_N\tau_3}{2M^2}} \vspace{0.5cm} \\
&& \frac{W+M_N}{8\pi W}\ \sqrt{\frac{E_f + M}{E_i+M}}\
\frac{W^2-M^2_N}{2W} \,, \nonumber\\
\lefteqn{\bar{M}_{1-}^{\rm FFR}  =  -\frac{eg_{\pi N}\kappa_N\tau_3}{2M^2}}
\vspace{0.5cm} \nonumber \\
&&\frac{W+M_N}{8\pi W}\ \frac{1}{\sqrt{(E_i+M)\,(E_f+M)}}\
\frac{W^2-M^2_N}{2W}\,. \nonumber
\end{eqnarray}
Both contributions are proportional to the photon energy $k$ that vanishes at
the nucleon pole position $W=M_N$. Reconstructing $A_1^{\rm FFR}$ from its
multipoles by Eq.~(\ref{eq:19}), one finds that the sum of the P-wave
contributions vanishes at the pole position, despite the denominators
$(W-M_N)^{-2}$ in that equation. As a consequence, only the S-wave contributes
to $A_1^{\rm FFR}$ at the nucleon poles.

While the tree diagrams can be calculated exactly, HBChPT provides the loop and
counterterm contributions as the first terms of a power series in the pion
energy $\omega$~\cite{Ber96,Ber01}. In particular, the loop contributions to
the S- and P-wave multipoles at threshold, $\omega=M_\pi$, are given by
\begin{eqnarray}
\label{eq:loops_thr}
\lefteqn{E_{0+}^{\rm loop} (\omega_{{\rm thr}})  =  {eg_A M_\pi^2\over 128 \pi
F_\pi^3}
 +{eg_A M_\pi^3\over 72 \pi^3 F_\pi^3 M_N}} \hspace{0.3cm}
\nonumber\\
&& \biggl\{-1-\frac{45\pi^2}{64}+\frac{33}{8} \,\ln\frac{M_\pi}{\lambda}
+\tau_3\biggl(-\frac{5}{4}+\frac{3}{2}\,\ln\frac{M_\pi}{\lambda}\biggr)\biggr\}
\nonumber\\
& &- {eg_A^3 M_\pi^3\over 512 \pi^3 F_\pi^3 M_N}
\biggl(\frac{44}{9}-\frac{20}{3}\pi+ \pi^2
-\frac{32}{3}\ln\frac{M_\pi}{\lambda}\biggr),
\\
&& \nonumber\\
\lefteqn{\bar P_1^{\rm loop}(\omega_{{\rm thr}}) = {eg_A^3 \, M_\pi\over 384
\pi^2 F_\pi^3}\left( 10   -
3\pi \right) + {eg_A \, M_\pi^2 \over 512 \pi^3F_\pi^3 M_N} } \hspace{0.3cm} \nonumber \\
&& \left\{ 16 +\pi^2+24 \ln{M_\pi \over \lambda}  +\tilde c_4 \left[{32\over 9}
+{64\over 3} \ln{M_\pi \over \lambda}\right]\right\}\nonumber\\
 & &
+{e g_A^3 \, M_\pi^2 \over 2304 \pi^3 F_\pi^3 M_N}\biggl\{20 - 2(\kappa_n
-\kappa_p) - 30\pi + 9\pi^2 \biggr. \nonumber \\ && + 6\tau_3
(1+\kappa_n+\kappa_p) \nonumber \\ && +12[4-\kappa_n+\kappa_p+3
\tau_3(1+\kappa_n +\kappa_p)]\ln{M_\pi \over \lambda} \biggr\}\,,
\label{eq:loop1_thr} \\
&& \nonumber\\
\lefteqn{\bar P_2^{\rm loop}(\omega_{{\rm thr}}) =
-{e g_A^3\, M_\pi \over 192\,\pi^2 F_\pi^3 }
-{e g_A\, M_\pi^2 \over 256\,\pi^3 F_\pi^3 M_N }} \hspace{0.3cm} \nonumber\\
& & \biggl\{-4+(\pi^2-\frac{80}{9})2\tilde c_4+\pi^2+
\bigr(8+\frac{16}{3}\tilde c_4\bigr)\ln\frac{M_\pi}{\lambda}\biggr\}
\nonumber\\
& & +{e g_A^3\, M_\pi^2 \over 192\,\pi^3 F_\pi^3 M_N }
\biggl\{\frac{\pi}{2}-\frac{3}{8}\pi^2+
\frac{11}{6}+\frac{2}{3}\kappa_n+\frac{1}{3}\kappa_p \nonumber\\
& & -\frac{1}{2}
(1+\tau_3)(1+\kappa_n+\kappa_p) - \bigl[4-2\kappa_p-4\kappa_n \nonumber\\
& &+3(1+\tau_3)
\bigl(1+\kappa_n+\kappa_p\bigr)\bigr]\ln\frac{M_\pi}{\lambda}\biggr\},
\label{eq:loop2_thr}\\
& &\nonumber \\
\lefteqn{\bar P_3^{\rm loop}(\omega_{{\rm thr}}) =
 -{e g_A\,M_\pi^2\over 256\pi F_\pi^3\,M_N }
(1+4\tilde c_4 ) }\hspace{0.3cm} \nonumber\\
& & -{e g_A^3\,M_\pi^2\over 192\,\pi^2 F_\pi^3\,M_N }
\{2(1-\kappa_n+\kappa_p)+\tau_3(1+\kappa_n+\kappa_p)\}\,,\nonumber\\
&&
\label{eq:loop3_thr}
\end{eqnarray}
where the low energy constant $\tilde{c}_4=M_Nc_4$ with $c_4=3.4$~GeV$^{-1}$.
The scale of dimensional regularization is set equal to the nucleon mass,
$\lambda=M_N$. Furthermore the axial coupling constant $g_A$ is fixed through
the Goldberger-Treiman relation, $g_A=g_{\pi N}F_\pi/M_N$ with $g_{\pi N}=13.1$
and $F_\pi=92.4$~MeV.

The counterterm contributions at threshold take the form~\cite{Ber96,Ber01}
\begin{eqnarray}
\label{eq:E0+_thr}
E_{0+}^{\rm ct}(\omega_{\rm{thr}}) &=& e\, (a_1^{p,n}(\lambda) + a_2^{p,n}
(\lambda))\,M_\pi^3\,,
\\
& &\nonumber \\ \label{eq:P2_thr} \bar P_{1,2}^{\rm ct}(\omega_{\rm {thr}}) &=&
{e g_A \, M_\pi^2 \over 64\pi^3 F_\pi^3 M_N} \, \xi_{1,2}^{p,n}(\lambda) \,,
\\
& & \nonumber \\ \label{eq:P3_thr}\bar P_3^{\rm ct}(\omega_{{\rm thr}})&=&
e \,b_P^{p,n}\left\{M_\pi -\frac{M_\pi^2}{2M_N}\right\}\,\ ,
\end{eqnarray}
where the low-energy constants have been either fitted to the threshold data or
estimated as the sum of vector meson exchange and $\Delta$ resonance
contributions by use of the resonance saturation principle (see Appendix~B).

By use of Eq.~(\ref{eq:a1_nonpole}), the dispersion relation of
Eq.~(\ref{eq:dr1}) can be cast into the form
\begin{eqnarray}
&&\kappa_N\tau_3 + \frac{2M_N^2}{eg_{\pi N}} \left\{
A_1^{\rm{loop}}(\nu,t_{\rm{thr}}) + A_1^{\rm{ct}}(\nu,t_{\rm{thr}})\right \}\nonumber\\
&=&\frac{4M_N^2}{\pi e\, g_{\pi N}}\,{\mathcal P}\int_{\nu_{\rm{thr}}}^\infty
d\nu' \frac{\nu'{\rm Im} A_1^{(N,\pi^0)}(\nu',t_{\rm{thr}})}{\nu'^2-\nu^2}\ ,
\label{eq:isoscalar}
\end{eqnarray}
If we insert Eqs.~(\ref{eq:loops_thr})-(\ref{eq:P3_thr}) into
Eq.~(\ref{eq:a1_thr}), we find immediately that the loop and counterterm
contributions in the curly bracket of Eq.~(\ref{eq:isoscalar}) vanish in the
soft-pion limit, $M_\pi\to 0$, which also leads to $\nu_{\rm{thr}}\to 0$ and
$t_{\rm{thr}}\to 0$. This result is in agreement with a low-energy theorem
derived on the basis of PCAC~\cite{Adl66} and has also been proved at the
one-loop order in relativistic ChPT~\cite{Ber92}. The result is the FFR sum
rule,
\begin{eqnarray}
\kappa_N\tau_3 = \frac{4M_N^2}{\pi\,eg_{\pi N}} \int_0^\infty d\nu'
\frac{{\rm{Im}}\,A_1^{(N,\pi^0)}(\nu',0)}{\nu'}\,.
\label{eq:FFRsum}
\end{eqnarray}

Of course, the imaginary part of the amplitudes in the dispersion integral
becomes a theoretical construct in the limit that describes the world of
massless pions. In order to stay in contact with the experimental data, our
strategy is to evaluate the RHS of Eq.~(\ref{eq:isoscalar}) along
$t=t_{\rm{thr}}$ and to study the loop and counterterm corrections as function
of $\nu$. Figure~\ref{mandelstam} shows that this is the only path $t=const$
for which ${\rm{Im}}\,A_1$ is directly related to the experimental data,
whereas all the other paths require an extrapolation into unphysical regions of
the Mandelstam plane. For further discussion we define the finite mass
corrections $\Delta_N$ to the FFR sum rule by the following equations:
\begin{eqnarray}
\lefteqn{\Delta_N(\nu,t_{\rm{thr}}) = } \vspace{0.3cm} \nonumber \\
&&  \frac{2M_N^2}{eg_{\pi N}} \left\{ A_1^{\rm{loop}}(\nu,t_{\rm{thr}}) +
A_1^{\rm{ct}}(\nu,t_{\rm{thr}})\right
\}\label{eq:FFR_Delta_N1}\,,\\
\lefteqn{\kappa_N\tau_3 + \Delta_N(\nu,t_{\rm{thr}})  =} \vspace{0.3cm}
\nonumber \\
&&  \frac{4M_N^2}{\pi e\, g_{\pi N}}\,{\mathcal P}\int_{\nu_{\rm{thr}}}^\infty
d\nu' \frac{\nu'{\rm Im} A_1^{(N,\pi^0)}(\nu',t_{\rm{thr}})}{\nu'^2-\nu^2}\,.
\label{eq:FFR_Delta_N}
\end{eqnarray}
We recall that the lower limit $\nu_{\rm{thr}}$ of the dispersion integral is
the threshold for neutral pion photoproduction. In practice, however, the
imaginary parts of the amplitudes are negligible below the onset of charged
pion production, which yields a strong cusp effect due to the large imaginary
part of the S-wave amplitudes.

\section{Results \label{sec:results}}

In order to obtain the correction $\Delta_N(\nu,t_{\rm{thr}})$ due to the
physical pion mass, we evaluate the invariant amplitude $A_1^{(N,\pi^0)}$ of
Eq.~(\ref{eq:isoscalar}) with the multipoles as predicted by HBChPT, whereas
the dispersion integral of Eq.~(\ref{eq:isoscalar}) is evaluated with the
MAID03 result for the imaginary part of the $A_1^{(N,\pi^0)}$ amplitude.

In Table~\ref{table1} we collect the results for the FFR sum rule of the proton
and neutron channels. The first column shows the experimental values of
$\kappa_p$ and $-\kappa_n$, the second column includes the finite mass
corrections at pion threshold as calculated from the loop and counterterms of
HBChPT (Eq.~(\ref{eq:isoscalar})). The results obtained from the dispersion
integral (RHS of Eq.~(\ref{eq:isoscalar})) are shown for $\nu=\nu_{{\rm{thr}}}$
in the third column and for $\nu=0$ in the last column.


\begin{table*}
\begin{center}
\begin{tabular}{|c|c|c|c|c|c|}
\hline\hline
  & FFR  & HBChPT ($\nu=\nu_{\rm{thr}}$)
& DR ($\nu=\nu_{\rm{thr}}$) & DR ($\nu=0$)
\\
\hline
& & & & \\
proton&
    1.793&
     2.29/2.33/2.37&
     2.24&
     1.66
     \\
& & & & \\
\hline
& & & & \\
neutron&
   1.913   &
   2.52/2.56/2.79 &
   2.44&
   1.82     \\
& & & & \\
\hline\hline
\end{tabular}
\end{center}
\caption{ The FFR sum rule for proton and neutron. First column: sum rule
values in the limit of $M_\pi\rightarrow 0.$ Second column: LHS of
Eq.~(\ref{eq:isoscalar}) as obtained from the HBChPT predictions for the S- and
P-wave multipoles evaluated at $\nu=\nu_{\rm{thr}}$. The results are given with
the LECs listed in Appendix~B according to  Refs.~\cite{Ber96}, \cite{BKM96},
and \cite{Ber01}, in order. Third column: dispersion prediction for the RHS of
Eq.~(\ref{eq:isoscalar}) at $\nu =\nu_{{\rm thr}}$ evaluated with MAID03.
Fourth column: dispersion prediction for the RHS of Eq.~(\ref{eq:isoscalar}) at
$\nu=0$ evaluated with MAID03.
\label{table1} }
\end{table*}

Both the HBChPT and the dispersion predictions yield large finite-mass
corrections at threshold. In the case of the proton we find
$\Delta_p(\nu_{\rm{thr}},t_{\rm{thr}})\approx 0.5$, in good agreement between
HBChPT and DR. The corresponding corrections for the neutron turn out to be
somewhat larger and take values of about 0.65. However it turns out that the
corrections $\Delta_N$ have become small and negative for $\nu=0$.

Table~\ref{table2} compares the experimental threshold values for the lowest
multipoles to the results of several theoretical descriptions. The latter
include the phenomenological model MAID03~\cite{Dre99}, the dispersion analysis
HDT~\cite{Han98}, and the dynamical (Dubna-Mainz-Taipei) model~\cite{Kam01}.
Concerning the comparison with the predictions of HBChPT, we note that we have
used the charged pion mass in all the loops and counterterms, and evaluated the
LECs with the $\Delta(1232)$ parameters given in Appendix~B. Occasionally, this
leads to a slight difference between the numbers in Tab.~\ref{table2} and those
given in Refs.~\cite{Ber96,BKM96,Ber01}, which however is irrelevant for our
further discussion. In the case of $E_{0+}^p$ we find a good agreement between
HBChPT and the phenomenological models, however there are some deviations for
the proton P-wave multipoles. Concerning the higher partial waves, only a
particular combination of the D waves contributes (see Eq.~(\ref{eq:a1_thr})),
and according to Table~\ref{table2} the non-pole contribution of these D waves
is small. We clearly observe somewhat larger differences among the predictions
for the neutron.

\begin{table*}
\begin{center}
\begin{tabular}{|l|c|c|c||c|}
\hline\hline & pole & MAID03/HDT/DMT & HBChPT & experiment\\
\hline &&&& \\
$E_{0+}^p$    & $-7.89$   & $-1.27/-1.22/-1.16$     & $-1.23/-1.15/-1.12$  & $-1.23\pm 0.08\pm0.03$ \\
$\bar{P}_1^p$ & $8.74$    & $9.35/9.64/9.31$        & $9.35/9.12/8.37$     & $ 9.46\pm 0.05\pm0.28$ \\
$\bar{P}_2^p$ & $-8.51$   & $-10.87/-10.49/-10.15$  & $-9.88/-9.61/-9.63$  & $ -9.5\pm 0.09\pm0.28$ \\
$\bar{P}_3^p$ & $0.59$    & $7.43/9.38/9.23$        & $12.90/10.63/5.90$   & $11.32\pm 0.11\pm0.34$ \\
$\bar{D}^p$   & $1.00$    & $0.96/$ - $/0.92$       & - & - \\
&&&& \\
\hline
&&&& \\
$E_{0+}^n$    & $-5.44$   & $1.47/1.19/1.93$        & $2.04/2.04/1.37$     & - \\
$\bar{P}_1^n$ & $5.98$    & $7.15/7.11/7.18$        & $6.66/6.50/6.64$     & - \\
$\bar{P}_2^n$ & $-5.98$   & $-8.71/-8.04/-8.15$     & $-7.46/-7.24/-7.87$  & - \\
$\bar{P}_3^n$ & $0.41$    & $7.05/8.80/8.46$        & $12.14/9.87/5.57$    & - \\
$\bar{D}^n$   & $-0.16$   & $-0.12/$ - $/-0.18$     & - & - \\
\hline\hline
\end{tabular}
\end{center}
\caption{ Threshold values of the S-, P-, and D-wave multipoles for neutral
pion photoproduction. Second column: pole contributions to the threshold
multipoles. Third column: results of the unitary isobar model
MAID03~\cite{Dre99}, the dispersion analysis HDT~\cite{Han98}, and the
dynamical model DMT~\cite{Kam01}, in order. Fourth column: the predictions of
HBChPT with the LECs given in Appendix~B according to Refs.~\cite{Ber96},
\cite{BKM96}, and \cite{Ber01}. The amplitudes $E_{0+}$ are in units of
$10^{-3}/M_{\pi^+}$, the P waves in units of $10^{-3}/M_{\pi^+}^2$, and the D
waves in units of $10^{-3}/M_{\pi^+}^3$. The experimental data are from
Ref.~\cite{Sch01}.
\label{table2}}
\end{table*}

\begin{figure}[ht]
\centerline{\resizebox{0.4\textwidth}{!}{%
  \includegraphics{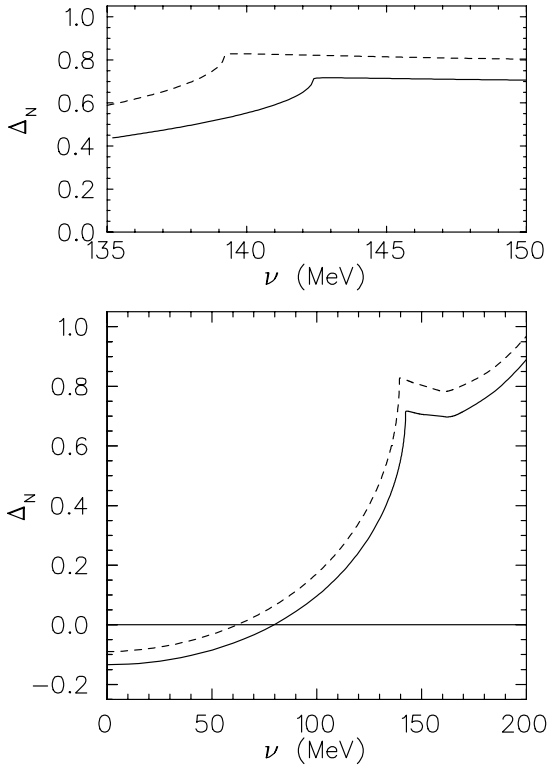}
}} \caption{ The correction to the FFR sum rule, $\Delta_N(\nu,t_{\rm{thr}})$,
as defined by Eq.~(\ref{eq:FFR_Delta_N}) and obtained from the dispersion
integral for proton (full line) and neutron (dashed line).}
\label{correction_pn}
\end{figure}

At this point we recall that the threshold of pion production moves to the
origin of the Mandelstam plane ($\nu=t=0$) if we approach the soft pion limit.
In order to get closer to this point we now study the dispersion integral of
Eq.~(\ref{eq:FFR_Delta_N}) as function of $\nu$ at $t=t_{\rm{thr}}$. This
integral is an even function of $\nu$, and therefore it must have an extremum
at $\nu=0$. It is also very likely that the value of the integral decreases
with decreasing $\nu$. However, the rapid decrease shown in
Fig.~\ref{correction_pn} is astounding.
The cusp effect at $\nu=\nu_{\rm{thr}}$ is very pronounced and leads to large
deviations from the sum rule. However, the importance of the loop effects
decreases rapidly if $\nu$ moves to values below threshold. The correction
vanishes at $\nu\approx70$~MeV, but because of the shallow minimum it is not
possible to predict the zero-crossing precisely.

In the following Fig.~\ref{integrand} we display the integrands of the
dispersion integrals for the isoscalar and isovector combinations. In the case
of the isovector combination (lower panel), the contribution of the
$\Delta(1232)$ is dominant. The cusp of threshold pion production is clearly
seen for $\nu=\nu_{\rm{thr}}$, however this effect reduces to a small shoulder
below the $\Delta$ resonance in the case of $\nu=0$. The contribution of the
second resonance region is small and practically independent of the choice of
$\nu$. The isoscalar combination (upper panel) takes, of course, much smaller
values than the isovector one. It also shows a strong cusp effect for
$\nu=\nu_{\rm{thr}}$, which rapidly decreases with smaller values of $\nu$.
However, the integrand peaks in the region of the $N^{\ast}(1520)$ and has some
additional strength of the opposite sign in the third resonance region.\\

\begin{figure}[ht]
\centerline{\epsfig{file=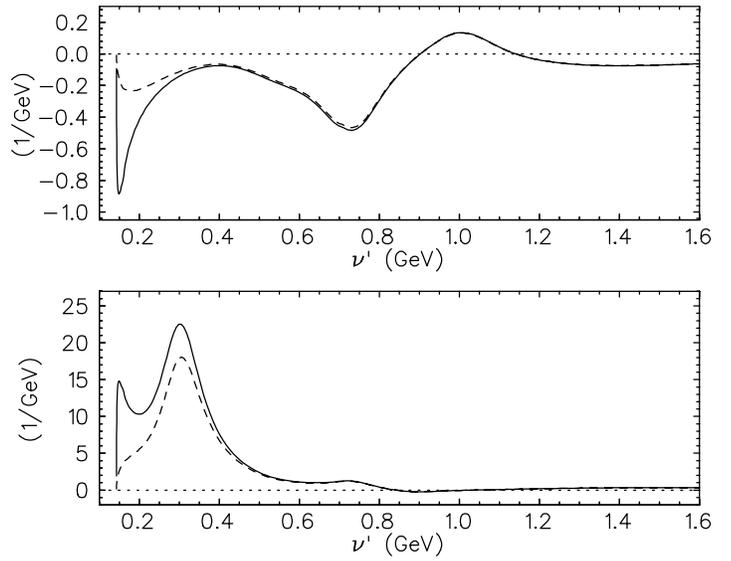, width=7.3cm, angle=90}}  \caption{The
integrands of the dispersion integrals (see RHS of Eq.~(\ref{eq:FFR_Delta_N}))
for the isoscalar (top) and isovector (bottom) combinations of the amplitudes
$A_1$. The full curves are obtained for $\nu=\nu_{\rm{thr}}$, the dashed curves
for $\nu=0$. }
\label{integrand}
\end{figure}

\begin{figure}[ht]
\begin{center}
\epsfig{file=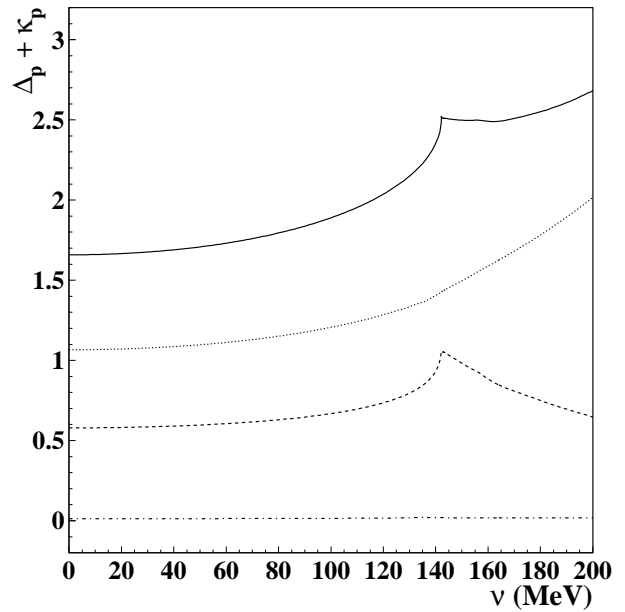, width=8cm}
\end{center}
\caption{The values of $\kappa_p+\Delta_p(\nu,t_{\rm{thr}})$ obtained from the
dispersion integral of Eq.~(\ref{eq:FFR_Delta_N}) with a multipole
decomposition of Im~$A_1$. Solid line: full result for Im$A_1$ evaluated with
MAID03. Dashed line: results for the dispersion integral with only the S-wave
contribution to Im$A_1$, dotted line: P-wave contribution only, dashed-dotted
line: sum of D- and F-wave contributions.\label{multipoles}}
\end{figure}

In Fig.~\ref{multipoles} we investigate the convergence of the multipole
expansion for the dispersion integral of the proton. The figure shows the
dispersion integral (RHS of Eq.~(\ref{eq:FFR_Delta_N})) evaluated over a large
energy range. The contribution to the integral value is clearly dominated by
the imaginary part of the P-wave amplitude, but the S-wave contribution is
substantial at low $\nu$ values and yields the cusp effect at threshold. The
imaginary parts of the higher partial waves turn out to be negligible over the
whole energy region.

At this point the reader should recall that the dispersion relation is valid
for the full invariant amplitudes $A_i$, not for the multipoles individually. A
decomposition of the real part of $A_i$ into a multipole series looks quite
different from Fig.~\ref{multipoles}, in that case the S wave provides the
lion's part of the amplitude in the region of $\nu\lesssim160$~MeV.

\begin{figure}[ht]
\begin{center}
\epsfig{file=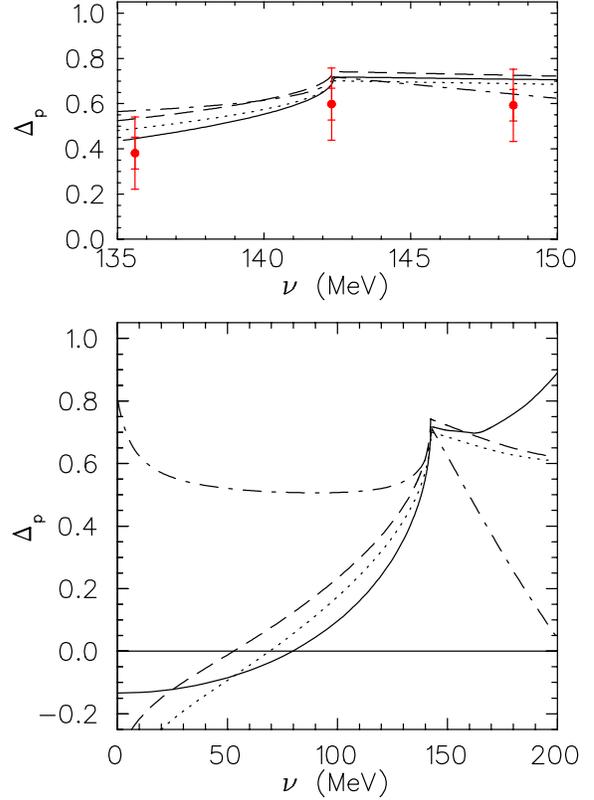, width=7.5cm }
\end{center} \caption{The correction to the FFR sum rule for the proton,
$\Delta_p(\nu,t_{\rm{thr}})$ as defined by Eqs.~(\ref{eq:FFR_Delta_N1}) and
(\ref{eq:FFR_Delta_N}). A comparison of the dispersive approach (solid lines)
with the predictions of HBChPT, represented by the dotted~\cite{Ber96},
dashed~\cite{BKM96}, and dashed-dotted~\cite{Ber01} lines. The data points are
calculated with the experimental S- and P-wave multipoles of Ref.~\cite{Sch01}.
\label{ffr_proton}}
\end{figure}

\begin{figure}[ht]
\begin{center}
\epsfig{file=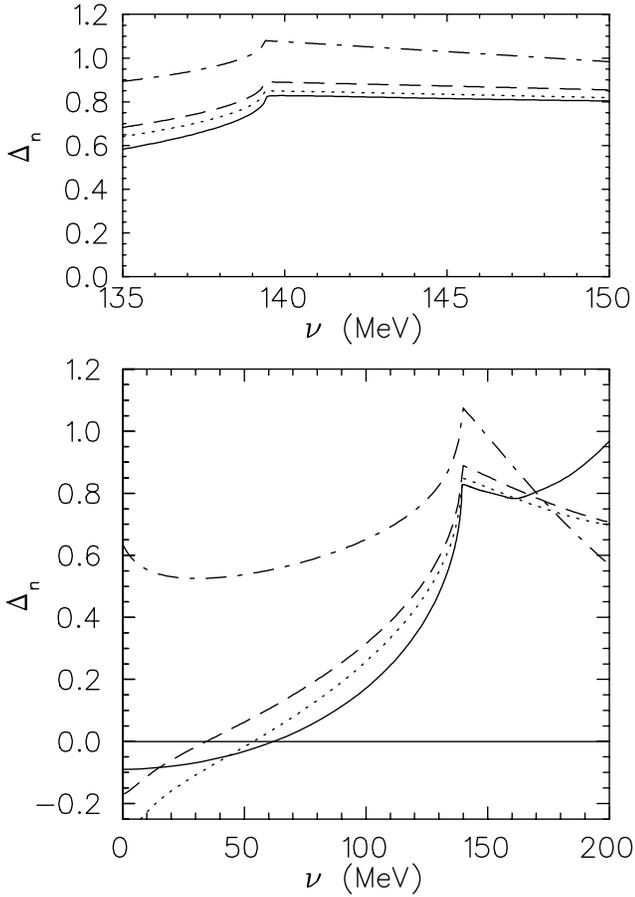, width=8.3cm }
\end{center} \caption{The correction to the FFR sum rule for the neutron,
$\Delta_n(\nu,t_{\rm{thr}})$ as defined by Eqs.~(\ref{eq:FFR_Delta_N1}) and
(\ref{eq:FFR_Delta_N}). A comparison of the dispersive approach (solid lines)
with the predictions of HBChPT, represented by the dotted~\cite{Ber96},
dashed~\cite{BKM96}, and dashed-dotted~\cite{Ber01} lines.
\label{ffr_neutron}}
\end{figure}

Figures~\ref{ffr_proton} and \ref{ffr_neutron} compare the predictions of
HBChPT and DR for $\Delta_N (\nu,t_{\rm{thr}})$ as functions of $\nu$ in the
range $0\le\nu\le200$~MeV, for the proton and the neutron, respectively. As we
have seen before, all the predictions agree quite well in the threshold region
(upper panels) , which includes the cusp effect due to the opening of the
charged pion channel.

There is also a reasonable agreement with the data points obtained by first
inserting the experimental values of the multipoles (Ref.~\cite{Sch01}, see
Table~\ref{table2}) in Eq.~(\ref{eq:19}) and then subtracting the pole terms.
As an example, the ``experimental'' threshold value has the following multipole
decomposition:
\begin{eqnarray}
\lefteqn{\kappa_p+\Delta_p(\nu_{\rm{thr}},t_{\rm{thr}})  =
2.06\,(S) + 0\,(\bar{P}_1) } \vspace{0.3cm} \\
&&  + 0.26\,(\bar{P}_2) - 0.19\,(\bar{P}_3) + 0.03\,(\bar{D}) = 2.16\,.
\nonumber
\end{eqnarray}
The result is clearly dominated by the S wave. However, the small total P-wave
term comes about by a delicate cancellation among the P waves, which leads to a
relatively large error bar for the FFR correction $\Delta_N$ if calculated from
the real part of the amplitude $A_1$. As we have seen in Figs.~\ref{integrand}
and \ref{multipoles}, the situation is quite different in the dispersive
approach. In this case, the correction $\Delta_N$ is essentially determined by
$\rm{Im}\,M_{1+}$ in the region of the $\Delta(1232)$ and a somewhat smaller
contribution of $\rm{Im}\,E_{0+}$ in the threshold region. Both contributions
are well under control and additive, and therefore the dispersive evaluation of
$\Delta_N$ should be quite stable.

Outside of the threshold region (lower panels), we observe 3 principal
differences between HBChPT and the dispersive approach:

\begin{enumerate}
\item[(I)]
The rise of $\Delta_N$ for $\nu\gtrsim170$~MeV is, of course, due to the
$\Delta(1232)$ resonance. It cannot be described by the ``static'' LECs of
Appendix~B, but will require a dynamical description of the resonance degrees
of freedom as, e.g., in the ``small scale expansion''~\cite{Hem97}. Such
effects could be approximately included by replacing the denominators
($\Delta^2-M_\pi^2$) in the resonance contributions of Appendix~B by
($\Delta^2-\omega^2$), which obviously leads to an enhancement of the resonance
effects for $\omega>M_\pi$.

\item [(II)] The curvature of the HBChPT predictions for small $\nu$ is due to the
nonrelativistic approach, which leads to (small) shifts of the nucleon pole
positions. As may be seen from Eq.~(\ref{eq:19}), the construction of the
relativistic amplitude $A_1$ requires that the multipole values be divided by
factors of $W-M_N$, which vanish at the position of the s-channel pole
$\nu=\nu_B=-9.7$~MeV. It is therefore unavoidable that nonrelativistic
approximations will lead to singularities at the s-channel pole, whereas the
u-channel pole at $\nu=-\nu_B=9.7$~MeV does not show up because of the angular
integration involved in the multipole expansion of Eq.~(\ref{eq:19}). As a
consequence nonrelativistic expansions are bound to yield large violations of
crossing symmetry in the region of small $\nu$ values.

\item [(III)] Except for the differences mentioned above, the predictions of
Refs.~\cite{Ber96} and \cite{BKM96} are in qualitative agreement with the
result of the dispersion approach as shown in Figs.~\ref{ffr_proton} and
\ref{ffr_neutron}. However, the prediction of Ref.~\cite{Ber01} leads to a
practically constant value of $\Delta_p\approx 0.55$ for $\nu<\nu_{\rm{thr}}$.
The reason for this difference is almost entirely due to the additionally
included ${\mathcal O}(q^4)$ loop terms in the P waves. Whereas this term is
essentially cancelled in $\bar{P}_2$ by a similarly large counterterm, the
cancellation is less effective for $\bar{P}_1$. As we note from
Eq.~(\ref{eq:a1_thr}), this fact does not show up near threshold where
$\bar{P}_1$ does not contribute to the amplitude $A_1$.

\end{enumerate}

The description of pion photoproduction by HBChPT was preceded by a
relativistic calculation at the one-loop order~\cite{Ber92}, which revised an
old low-energy theorem and  explained the strong suppression of neutral pion
photoproduction on the proton by loop effects. Unfortunately, it was then not
possible to define a systematic power counting underlying such a relativistic
field theory if nucleons were involved. This problem could be solved by HBChPT,
which organizes the Lagrangian as a power series in $1/M_N$ in order to obtain
a well-defined expansion of the observables if the typical external momenta of
the system are small compared to the nucleon mass. As we have seen above,
HBChPT can indeed well describe the amplitudes in the threshold region, whereas
it fails in the region of small $\nu$ values. Because of its non-relativistic
approximations, it misses the nucleon pole positions and thus fails to
reproduce the crossing-symmetry, which is one of the essentials of a dispersive
approach. The shortcomings of HBChPT have of course been noted often before,
and several groups are now working to apply the newly developed manifestly
Lorentz-invariant renormalization schemes~\cite{Bec99} to various physical
processes, in particular also to pion photoproduction. However, the general
structure of the dispersive part of the amplitude $A_1$ for small external
momenta has already been given in Ref.~\cite{Ber92}:
\begin{equation}
A_1^{\rm{disp}} = a_{00} + a_{02}\nu_B + a_{20}\nu^2 + \ldots\,,
\label{A1_disp}
\end{equation}
where the coefficients $a_{ik}$ are functions of the mass ratio
$\mu=M_\pi/M_N$. In particular the leading coefficients depend on the pion mass
as follows: $a_{00}={\mathcal O}\,(\mu^2)$, $a_{02}={\mathcal O}\,(\ln\,\mu)$,
and $a_{20}={\mathcal O}\,(\mu^{-1})$. The vanishing of $a_{00}$ in the chiral
limit is, of course, a necessary condition for the validity of the FFR sum
rule. Furthermore, the divergence of the higher expansion coefficients in that
limit is the reason why the old low-energy theorem for neutral pion
photoproduction failed.

\begin{figure}[ht]
\begin{center}
\epsfig{file=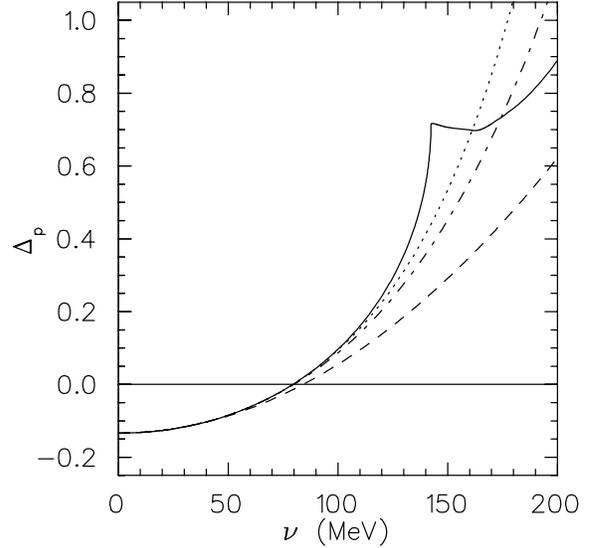, width=7.5cm }
\end{center} \caption{Full line: the correction $\Delta_p(\nu,t_{\rm{thr}})$ as defined
by Eq.~(\ref{eq:FFR_Delta_N}), compared to the low-energy expansion in the
crossing-symmetrical variable $\nu$. Dashed line: term of ${\mathcal O}
(\nu^2)$, dashed-dotted line: up to ${\mathcal O} (\nu^4)$, and dotted line: up
to ${\mathcal O} (\nu^6)$.
\label{LEX}}
\end{figure}

Since the FFR discrepancy $\Delta_N(\nu,t)$ is directly related to
$A_1^{\rm{disp}}$, it has a power series expansion similar to
Eq.~(\ref{A1_disp}),
\begin{equation}
\Delta_N(\nu,t) = \delta_{00}^N + \delta_{02}^N\,\frac{t-M_\pi^2}{4M_N} +
\delta_{20}^N\nu^2 + \ldots \,.
\label{DeltaN_nut}
\end{equation}
In particular $\delta_{00}$ can be determined by an analytical continuation of
the multipole expansion of Eq.~(\ref{eq:F1_4}) to the unphysical point
($s=u=M_N^2,\ t=M_\pi^2$) at which all particles are on their mass shell (see
Fig.~\ref{mandelstam}), i.e.,
\begin{equation}
\delta_{00}^N = \Delta_N(\nu=0,\ t=M_\pi^2) \,.
\label{delta00}
\end{equation}
This extrapolation requires some care, because the Legendre polynomials
$P_\ell(x)$ involved in the expansion have to be evaluated at $|x|>1$. However,
preliminary studies with only S- and P-wave contributions indicate that
$\delta_{00}^N$ is small. The remaining two constants in Eq.~(\ref{DeltaN_nut})
can be approximately obtained by DRs at $t=t_{\rm{thr}}$, in which case the
path of integration is fully contained in the physical region. The value of
$\delta_{02}$ can be read off at $\nu=0$,
\begin{equation}
\delta_{02}^N \approx
\frac{4M_N}{t_{\rm{thr}}-M_\pi^2}\,(\Delta_N(\nu=0,t_{\rm{thr}})-\delta_{00}^N)
 \,,
\label{delta02}
\end{equation}
where the ellipses denote the influence of higher order terms as in
Eq.~(\ref{DeltaN_nut}). The coefficients of $\nu^{2n}$, $n\ge1$, can be easily
obtained by expanding the dispersion integral of Eq.~(\ref{eq:FFR_Delta_N}),
e.g.,
\begin{equation}
\delta_{2n,0}^N \approx \frac{4M_N^2}{\pi e g_{\pi
N}}\,\int_{\nu_{\rm{thr}}}^\infty \frac{d\nu'}{(\nu')^{2n+1}}\,{\rm
Im}\,A_1^{(N,\pi^0)}(\nu',t_{\rm{thr}}) \,.
\label{delta20}
\end{equation}
Due to the additional factors of $1/\nu'^2$, these integrals are well saturated
by the threshold and $\Delta(1232)$ resonance regions. The numerical results
for these coefficients are $\delta_{20}^p=0.368/M^2_{\pi+}$,
$\delta_{40}^p=0.120/M^4_{\pi+}$, and $\delta_{60}^p=0.054/M^6_{\pi+}$, and
Fig.~\ref{LEX} shows the convergence of the respective Taylor series below pion
threshold.

\section{Summary and Outlook\label{sec:sum}}

The Furbini-Fulan-Rossetti (FFR) sum rule connects the nucleon's anomalous
magnetic moment with the neutral pion photoproduction amplitudes
$A_1^{(+,0)}(\nu,t)$ in the chiral limit of massless pions. We have studied the
finite mass corrections to this sum rule in the framework of dispersion
relations at $t=const$ as function of the crossing-symmetrical variable $\nu$.
The path of integration has been chosen such that it passes through the
threshold of pion photoproduction, $\nu=\nu_{\rm{thr}}$ and $t=t_{\rm{thr}}$,
because this is the only t-value for which the integrand of the DR is fully
contained in the physical region. A comparison with the results of heavy baryon
chiral perturbation theory (HBChPT) shows excellent agreement in the threshold
region but deviations for both higher and lower energies. For energies moving
towards the $\Delta(1232)$ resonance, the agreement can be improved by
including the dynamical degrees of freedom of that resonance as, e.g., in the
``small scale expansion''. For energies far below threshold, the problem lies
in the nonrelativistic approximations involved in HBChPT. As a consequence of
these approximations, the nucleon poles are situated at the wrong positions,
which leads to singularities in the dispersive amplitude and to a violation of
the crossing symmetry. The remedy for these problems lies in a manifestly
Lorentz-invariant treatment of pion photoproduction, which  at the same time
should also provide a systematic expansion scheme. This challenge has prompted
several groups to take up the issue, and a solution of the problem is to be
expected soon.

The aim of our work is two-fold. First, we want to put our phenomenological
model MAID on a dispersive basis, i.e., the real and imaginary parts of the
pion production amplitudes should be Hilbert transforms of each other. The
obvious technique to connect these amplitudes are DRs at $t=const$, which in
general requires an integration over physical and unphysical regions. The
present version of MAID03 involves all four-star S-, P-, D-, and F-wave
resonances up to total c.m. energy $W=2$~GeV, and the convergence problem of
the multipole series in the unphysical region has to be studied with great
care. In this sense we see the FFR sum rule as a good testing ground for the
development of a ``dispersion MAID''.

The second aim of our work is to connect the dispersive approach to the soon
expected results from relativistic ChPT. The low-energy constants (LECs) of
such a theory can be provided by dispersion integrals that contain global
properties of the resonance spectrum. Since these LECs are the coefficients of
a power series in small $\nu,\ t$, and $M_\pi$, the ideal meeting ground
between relativistic ChPT and DRs is in the unphysical region around the origin
of the Mandelstam plane, $\nu=t=0$.

In our work we have concentrated on the FFR sum rule and the amplitudes
$A^{(+,0)}$ for neutral pion photoproduction. It will be straightforward to
extend our approach to charged pion photoproduction and, eventually, to all 6
amplitudes of pion electroproduction. We hope that this future work will
provide many new cross-checks with relativistic effective field theories,
improve our understanding of pion production as the most prominent decay mode
of the nucleonic resonance spectrum, and lead to new insights concerning the
interplay between pion-cloud and resonance structure of the nucleon.

\section*{Acknowledgements}

The authors gratefully acknowledge several discussions with Norbert Kaiser and Ulf-G.
Mei{\ss}ner on various aspects of Chiral Perturbation Theory. This work was supported by
the Italian MIUR through the PRIN Theoretical Physics of the Nucleus and the Many-Body
Systems (B. Pasquini) and by the Deutsche Forschungsgemeinschaft (SFB 443). This research
is part of the EU Integrated Infrastructure Initiative Hadron Physics Project under
contract number RII3-CT-2004-506078.

\section*{Appendix}
\begin{appendix}
\section{Pole contribution}

The pole contribution to the S- and P-wave multipoles at threshold can be
derived from the invariant amplitudes given in Eq.~(\ref{eq:a1-4pole}),
\begin{eqnarray}
\lefteqn{E_{0+}(\omega_{thr})= \frac{\mu
M_N}{8\pi}\,\frac{2+\mu}{(1+\mu)^{3/2}}}  \vspace{0.8cm} \nonumber \\
&& \quad \left[A_1+ \mu M_N\,\frac{2+\mu}{2(1+\mu)}\, A_3
+ \frac{\mu^2\, M_N}{2(1+\mu)}\, A_4\right],\nonumber\\
\lefteqn{\bar{P}_2(\omega_{thr})=\frac{\mu
}{16\pi}\,\frac{(2+\mu)}{(1+\mu)^{3/2}}}\vspace{0.8cm} \nonumber \\
&& \quad \left[-A_1+ 2 \mu M_N^2\,A_2+ \frac{M_N(2+\mu)^2}{2(1+\mu)} \right.
\nonumber \\ && \quad \left. A_3
 + \mu M_N\, \frac{2+\mu}{2(1+\mu)}\, A_4\right],\nonumber\\
\lefteqn{\bar{P}_3(\omega_{thr})=\frac{\mu
}{16\pi}\,\frac{2+\mu}{(1+\mu)^{3/2}} } \vspace{0.8cm} \nonumber \\
&& \quad \left[-\,A_1+ \frac{\mu^2 M_N}{2(1+\mu)}\,A_3
 + M_N\, \frac{4+6\mu+\mu^2}{2(1+\mu)}\, A_4\right],\nonumber\\
\label{eq:A1}
\end{eqnarray}
where  $\mu=M_\pi/M_N$ and $A_i=A_i(s_{\rm{thr}},\ t_{\rm{thr}})$. The
threshold values of the latter amplitudes follow from Eq.~(\ref{eq:a1-4pole}):
\begin{eqnarray}
A_1^{{\rm pole}}&=&-\frac{1+\tau_3}{2}\frac{eg_{\pi N}}{M_N^2}
\frac{1}{(2+\mu)}\, ,
\nonumber\\
A_2^{{\rm pole}}&=&-(1+\tau_3)\frac{eg_{\pi N}}{M_N^4}
\frac{(1+\mu)}{\mu^2(2+\mu)^2}\, ,
\nonumber\\
A_3^{{\rm pole}}&=&-\tau_3\,\kappa_N\frac{eg_{\pi N}}{2M_N^3\mu}\, ,\nonumber\\
A_4^{{\rm pole}}&=&
\tau_3\,\kappa_N\frac{eg_{\pi N}}{2M_N^3}\frac{1}{(2+\mu)}\, .
\label{eq:A2}
\end{eqnarray}
The threshold values of the 3 multipoles given by Eqs.~(\ref{eq:A1}) can be
directly obtained by inserting the threshold amplitudes of Eqs.~(\ref{eq:A2}).
In the case of $\bar{P}_1^{{\rm pole}}(\omega_{\rm{thr}})$, we first have to
project out the P-wave content of the invariant amplitudes before going to the
threshold kinematics. As a result the pole contributions to the S- and P-wave
multipoles take the form
\begin{eqnarray}
E_{0+}^{{\rm pole}}(\omega_{thr})&=&-\frac{eg_{\pi N}}{8\pi M_N}
\frac{\mu}{(1+\mu)^{3/2}}\left[\frac{1+\tau_3}{2}+\tau_3\kappa_N\right]\,,
\nonumber\\
\bar{P}_1^{{\rm pole}}(\omega_{thr})&=&\frac{eg_{\pi N}}{16\pi M_N^2}
\frac{2+\mu}{(1+\mu)^{3/2}}\left[\frac{1+\tau_3}{2}
+\tau_3\kappa_N\right]\,,\nonumber\\
\bar{P}_2^{{\rm pole}}(\omega_{thr})&=&-\frac{eg_{\pi N}}{16\pi M_N^2}
\frac{2+\mu}{(1+\mu)^{3/2}} \nonumber \\ && \left[\frac{1+\tau_3}{2}\left(1
-\frac{2\mu(1+\mu)}{(2+\mu)^2}\right)+\tau_3\kappa_N\right]\,,\nonumber\\
\bar{P}_3^{{\rm pole}}(\omega_{thr})&=&\frac{eg_{\pi N}}{16\pi M_N^2}
\frac{\mu}{(1+\mu)^{3/2}}\left[\frac{1+\tau_3}{2}+\tau_3\kappa_N\right] \, .
\nonumber \\
&&
\label{eq:A3}
\end{eqnarray}
The FFR contributions are obtained by evaluating Eq.~(\ref{eq:E0+_M1-}) at
threshold:
\begin{eqnarray}
\label{eq:A4}
E_{0+}^{{\rm FFR}}(\omega_{thr}) &=& \frac{eg_{\pi N}\,\tau_3\kappa_N}{16\pi
M_N} \frac{\mu(2+\mu)}{(1+\mu)^{3/2}}\,,  \\
\bar{P}_1^{{\rm FFR}}(\omega_{thr}) &=& -\bar{P}_2^{{\rm
FFR}}(\omega_{thr})=-\bar{P}_3^{{\rm FFR}}(\omega_{thr}) \nonumber \\ & = &
\frac{eg_{\pi N}\,\tau_3\kappa_N}{32\pi M_N^2}
\frac{\mu(2+\mu)}{(1+\mu)^{3/2}}\ .\nonumber
\end{eqnarray}
The sum of Eqs.~(\ref{eq:A3}) and (\ref{eq:A4}) yields the result of
pseudo-vector pion-nucleon coupling, which agrees with the expansion of
Refs.~\cite{Ber96,Ber01} up to ${\mathcal{O}}(1/M_N^4)$.


\section{Low energy constants of HBChPT}

In the following we list the low-energy constants determined by several
investigations on neutral pion photoproduction off the proton.

\begin{enumerate}
\item [(I)]Resonance fit of Ref.~\cite{Ber96}:

The low-energy constants (LECs) are obtained by the resonance saturation
principle. The t-channel exchange of the vector mesons $\rho(770)$ and
$\omega(782)$ leads to:
\begin{eqnarray}
a_1^V & = & -\frac{1}{24\pi^3M_NF_\pi^3}\,, \nonumber \\
a_2^V & = & \frac{5}{48\pi^3M_NF_\pi^3}\,,\nonumber \\
b_P^V & = & \frac{5}{64\pi^3F_\pi^3}\,,
\end{eqnarray}
where $F_\pi=93$~MeV has been used. The largest resonance contribution in the
s-channel is given by $\Delta(1232)$ excitation. The corresponding contribution
$a_1^\Delta,\ a_2^\Delta$ and $b_P^\Delta$ were determined according to
Eqs.~(4.8)-(4.9) of Ref.~\cite{Ber96} with the following parameters:
$C=0.40$~GeV$^{-5}$, $g_1=g_2=5$, $X=2.24$, $Y=0.13$, $Z=0.28$. The resulting
values are: $a_1^\Delta=1.26$~GeV$^{-4}$, $a_2^\Delta=2.62$~GeV$^{-4}$,
$b_P^\Delta=12.75$~GeV$^{-3}$. The LECs of model (I) were then obtained by
adding the s- and t-channel contributions: $a_1=-0.52$~GeV$^{-4}$,
$a_2=7.07$~GeV$^{-4}$, $b_P=15.88$~GeV$^{-3}$.

\item[(II)] Resonance fit of Ref.~\cite{BKM96}:

The LECs follow from the same procedure as above except for different values of the
off-shell parameters of $\Delta$ excitation, $X=2.75$, $Y=0.10$, $Z=-0.21$. The result is
$a_1=(-1.78+2.46)$~GeV$^{-4}=0.68$~GeV$^{-4}$,
$a_2=(4.45+1.45)$~GeV$^{-4}=5.90$~GeV$^{-4}$,
$b_P=(3.13+9.89)$~GeV$^{-3}=13.03$~GeV$^{-3}$, where the first values in the brackets
refer to the vector meson, the second value to the $\Delta$ contribution.

\item[(III)] Five-parameter fit of Ref.~\cite{Ber01}:

This calculation includes both S- and P-waves consistently to fourth order in
HBChPT. The S-wave LECs are taken from a comparison with differential cross
sections and photon asymmetries (Set II of Table~\ref{table2} in
Ref.~\cite{Ber01}), $a_1=8.588$~GeV$^{-4}$, $a_2=-2.288$~GeV$^{-4}$, while the
3 P-wave LECs follow from resonance saturation,
\begin{equation}
b_P^V=\frac{5}{64\pi^3F^3_\pi}\,,\quad \xi_1^V=-\frac{8}{g_A}\,,\quad
\xi_2^V=\frac{4}{g_A}\,,
\end{equation}
with $F_\pi=92.4$~MeV and $g_A=g_{\pi N}F_\pi/M=1.29$. The $\Delta$
contributions are determined from
\begin{eqnarray}
b_P^\Delta  & = & \frac{\kappa^{\ast}g_A}{6\sqrt{2}\pi
M_NF_\pi}\,\frac{\Delta}{\Delta^2-M_\pi^2} \,, \nonumber \\
\xi_1^\Delta & = & -\xi_2^\Delta
=\frac{\kappa^{\ast}}{3\sqrt{2}}\,\frac{16\pi^2 F_\pi^2}{\Delta^2-M_\pi^2}\,,
\end{eqnarray}
where $\Delta$ is the $N\Delta$ mass splitting and $\kappa^*$ the $N\Delta$
transition magnetic moment. With $\Delta=293$~MeV and $\kappa^*=4.86$, the
total result for the LECs is
\begin{eqnarray}
b_P & = & (3.19+11.73)\ {\rm{GeV}}^{-3} =
14.93\ {\rm{GeV}}^{-3}\,, \nonumber\\
 \xi_1& = & -6.21+22.84=16.63\,, \nonumber \\
 \xi_2 & = & 3.10-22.84 = -19.73\,.
\end{eqnarray}
The LECs for the neutron differ from those of the proton in the vector meson
contributions according to
\begin{equation}
a_i^{V,n} = \frac{1}{8}a_i^{V,p}\,,\quad b_P^{V,n} = \frac{4}{5}
b_P^{V,p}\,,\quad \xi_1^{V,n}= \frac{1}{8}\xi_i^{V,p}\,,
\end{equation}
while the $\Delta$ resonance contributions remain unchanged.

\end{enumerate}

\end{appendix}

\end{document}